
\magnification=1200
\hsize=15truecm
\vsize=23truecm
\baselineskip 20 truept
\voffset=-0.5truecm
\parindent=0cm
\overfullrule=0pt
\null

\centerline{\bf Light--like Wilson loops and gauge}
\centerline{\bf invariance of Yang--Mills theory}
\centerline{\bf in 1+1 dimensions}
\vskip 1.0truecm
\centerline{A. Bassetto}
\centerline{\it Dipartimento di Fisica ``G. Galilei'',
Via Marzolo 8 - 35131 Padova, Italy}
\centerline{\it INFN, Sezione di Padova, Italy}

\vskip 0.5truecm
\centerline{F. De Biasio}
\centerline{\it Dipartimento di Fisica ``G. Galilei'',
Via Marzolo 8 - 35131 Padova, Italy}
\vskip 0.5truecm

\centerline{L. Griguolo}
\centerline{\it SISSA, Via Beirut 2 - 34100 Trieste, Italy}
\centerline{\it INFN, Sezione di Trieste, Italy}
\vskip 1.0 truecm
\centerline{\bf Abstract}
\vskip 0.3truecm
A light-like Wilson loop is computed in perturbation theory
up to ${\cal O} (g^4)$ for pure Yang--Mills theory in 1+1 dimensions,
using Feynman and light--cone
gauges to check its gauge invariance. After dimensional regularization in
intermediate steps, a finite gauge invariant result is obtained, which
however does not exhibit abelian exponentiation. Our result is at variance
with the common belief that pure Yang--Mills theory is free in 1+1
dimensions, apart perhaps from topological effects.

\vskip 0.3truecm
DFPD 94/TH/2 \hskip 5truecm January 1994.
\vfill\eject

{\bf 1. Introduction}
\vskip 0.3truecm
While an abundant literature  is by now available
concerning ``euclidean" QCD in two dimensions [1], [2], [3]
comparatively fewer investigations have been performed in minkowskian
1+1 dimensions.

It is a common belief, taking rise from the pioneeristic work of
't Hooft [4], that Yang--Mills theory (YMT) without fermions
in two dimensions is a free theory apart perhaps from topological
effects. As a matter of fact, the gauge field should not be
endowed with any dynamical degree of freedom.
This can naively be seen in axial gauges, where one of
the components
of the vector potential is set equal to zero. It is also at the
root of the possibility of calculating the mesonic spectrum in the
large N approximation [4], [5], when quarks are introduced.

However  the theory exhibits severe infrared (IR) divergences which
need to be regularized. In ref. [4] an explicit IR cutoff is advocated,
which turns out to be uninfluent in the bound state equation; a
Cauchy principal value (CPV) prescription in handling such IR singularity
leads indeed to the same result [6]. Still difficulties in performing a
Wick's rotation in those conditions have been pointed out [7], and
a causal prescription for the IR singularity has been advocated,
leading to a quite different solution for the vector propagator.
In this context the bound state equation with vanishing bare quark
masses [8] has solutions with quite different properties, when
compared with the ones of refs.[4],[5].

In view of the above mentioned controversial results and of the
fact that ``pure" YMT does not immediately look free in Feynman gauge,
where degrees of freedom of a ``ghost"-type are present,
we have thought worth performing a test on a gauge invariant
quantity. We have chosen a rectangular Wilson loop with
light--like sides, directed along the vectors $n_\mu = (T, - T)$ and
$n^*_\mu = (L, L)$, parametrized according to the equations:

$$
\eqalign{
C_1 : & x^\mu (t) = n^{* \mu} t,\cr
C_2 : & x^\mu (t) = n^{* \mu} + n^\mu t,\cr
C_3 : & x^\mu (t) = n^\mu + n^{* \mu}( 1-t), \cr
C_4 : & x^\mu (t) = n^\mu (1 - t), \qquad 0 \leq t \leq 1. \cr}
\eqno(1.1)
$$

This contour has been considered in refs. [9],[10] for
an analogous test of gauge invariance in 1+3 dimensions. Its
light--like character forces a Minkowski treatment.

We shall perform a perturbative calculation up to
${\cal O}(g^4)$; in so doing
topological effects will not be considered. We can anticipate
the unexpected results we will obtain: the gauge invariant
theory is not free at d= 1+1, at variance with the
commonly accepted behaviour; the theory in d= 1+(D -- 1) is
``discontinous" in the limit $D \rightarrow 2$.
\vskip 0.3truecm
{\bf 2. The calculation in Feynman gauge}
\vskip 0.3truecm
In Feynman gauge already the free vector propagator
does not exist as a tempered distribution in 1+1
dimensions. A regularization is thereby mandatory
and we choose from here on to adopt the dimensional
regularization, which preserves gauge invariance:

$$
D^F_{\mu\nu} (x) = - g_{\mu\nu} {\pi^{-D/2} \over 4}
\Gamma (D/2 -1) (- x^2 + i \epsilon)^{1 - D/2},
\eqno(2.1)
$$

The calculation in Feynman gauge of the light--like
Wilson loop (1.1), in 1+(D -- 1) dimensions up to
${\cal O}(g^4)$ has been performed in ref. [9] and will
not be repeated here. Actually in ref. [9] part of the
contribution from graphs containing three vector
lines, has only been given as a Laurent expansion
around D=4, owing to its complexity. In the following
we shall exhibit its general expression in terms of
a generalized hypergeometric series and then we shall
expand it around D=2.
We only report for the reader's benefit the final results
concerning the contributions of the various diagrams.

The single vector exchange $({\cal O}(g^2)$) gives

$$
W_F^{(1)} = - \Bigl({\hat g \over \pi\mu}\Bigr)^2 C_F {\Gamma(D/2 - 1)
\over (D - 4)^2} {\cal C}, \eqno (2.2)
$$

where $\hat g^2$ = $g^2 (\mu^2)^{D/2 -1}$, $\mu$ being the
running renormalization mass, $C_F$ is the  Casimir
operator of the fundamental representation of color ${\it su}$(N) and

$$
{\cal C} \equiv \Bigl[( 2\pi \mu^2 nn^* + i \epsilon)^{2 -D/2}
+ (- 2\pi \mu^2 nn^* + i \epsilon)^{2-D/2}\Bigr]. \eqno(2.3)
$$

One immediately notice that the propagator pole at D=2 is
cancelled after integration over the contour, leading to a finite result.

At order ${\cal O}(g^4)$ we can restrict ourselves to the
so--called ``maximally non--abelian" contributions [11].
The self--energy correction to the propagator gives

$$
W_F^{(2 ; se)} = ({\hat g \over \pi\mu})^4 {C_F C_A \over
64} {\Gamma^2 (D/2 - 1) (3D-2) \over (D-4)^3 (D-3)(D-1)}
{\cal A}, \eqno(2.4)
$$

where $C_A$ is the Casimir operator of the adjoint representation,

$$
{\cal A} \equiv [(2 \pi \mu^2 nn^* + i \epsilon)^{4-D}
+(- 2\pi \mu^2 nn^* + i \epsilon)^{4-D}] \eqno(2.5)
$$

and the fermionic loop has not been considered
(pure YMT). Eq. (2.4) exhibits a double pole at D=2.

Next we consider the contribution of the so--called ``cross"
graphs, the ones with two non interacting crossed vector
exchanges

$$
W_F^{(2;cr)} = - ({\hat g \over \pi\mu})^4
{C_F C_A \over 16} {\Gamma^2 (D/2 - 1) \over (D - 4)^4}
[{\cal A}+ 8 {\cal B} (1 - {\Gamma^2 (3 -D/2) \over \Gamma(5-D)})],
\eqno(2.6)
$$

where
$$
{\cal B} \equiv [( 2\pi \mu^2 nn^* + i \epsilon)
(-2 \pi \mu^2 nn^* + i \epsilon)]^{2 - D/2}. \eqno(2.7)
$$

Again a double pole occurs at D=2.

The contribution coming from graphs with three vector lines
is by far the most complex one. It is convenient to split
it into two parts, one coming from graphs with two vector lines
attached to the same side

$$
W_F^{(2; ss)} = ({\hat g \over \pi \mu})^4 {C_F C_A \over 16}
{{\cal A} \over (D - 4)^4} [2 \Gamma (3 - D/2) \Gamma (D/2 - 1)
\Gamma (D - 3) - {1 \over D - 3} \Gamma^2 ( D/2 - 1)]
\eqno (2.8)
$$
and another one in which the three ``gluons" end in three
different rectangle sides

$$
W_F^{(2; ds)} = \Bigl({\hat g \over \pi \mu}\Bigr)^4 {C_F C_A \over 64}
{\cal A} \Bigl\{{\Gamma^2 (D/2 - 2) \Gamma (4 - D/2) \Gamma
(D - 3) \over \Gamma (D/2)} F (D)+
$$
$$
+ {4 \over (D - 4)^4}
\Bigl[\Gamma (3 - D/2) \Gamma (D/2 - 1) \Gamma (D - 3)
- \Gamma^2 (D/2 - 1)\Bigr]\Bigr\}. \eqno(2.9)
$$

The function F (D) is defined as

$$
F(D) = S(D) + {D/2 - 1 \over (3 - D/2) (D - 4)} [5 \psi (3 - D/2)
- \psi (D/2 - 1) - 2 \psi (1)
$$
$$
- 2 \psi (5 - D)], \eqno(2.10)
$$

$\psi (D)$ being the digamma function and S(D) the convergent
generalized hypergeometric series

$$
S(D) = \sum^\infty_{n=0} {1 \over (n + 1)^2} {1 \over n!}
{\Gamma (n + D - 3) \over \Gamma (D - 3)} {\Gamma (n + 4 - D/2) \over
\Gamma ( 4 - D/2)} {\Gamma (D/2) \over \Gamma (n + D/2)}. \eqno(2.11)
$$

Both contributions exhibit a double pole at D=2.
The Laurent expansion of eq. (2.9) around D=4, reproduces exactly
the expression given in ref. [9].

Summing eqs. (2.4), (2.6), (2.8) and (2.9) and performing a
careful Laurent expansion around D=2, it is tedious but
straightforward to prove that double and single poles cancel,
leaving  only the finite contribution

$$
W_F^{(2)} (D = 2) = ({g^2 \over 4 \pi})^2 C_F C_A (n^* n)^2
(1 + {\pi^2 \over 3}). \eqno (2.12)
$$

The presence of a non vanishing $C_F C_A$ contribution is
a dramatic result: it means that the theory does not
exponentiate in an abelian way, as a ``bona fide" free theory
should do. In order to better understand this result, it is worth
turning now our attention to the same Wilson loop calculation,
performed in the light--cone axial gauge nA = 0.
\vskip 0.3truecm
{\bf 3. The calculation in light--cone gauge nA = 0}
\vskip 0.3truecm
The free vector propagator in light--cone gauge
is very sensitive to the prescription used to
handle the so--called ``spurious" singularity.
The only prescription, known so far, which allows
to perform a Wick's rotation without extra terms
and to calculate loop diagrams in a consistent way
[12] is the causal Mandelstam--Leibbrandt (ML) prescription
[13]. In a canonical formalism it is obtained by
imposing equal time commutation relations [14];
in two dimensions a ``ghost" degree of freedom still
survives, as will be discussed in the last section.

When ML prescription is adopted, the free vector
propagator is indeed a tempered distribution at
D=2 [15], at variance with its behaviour in
Feynman gauge. In particular, when $x_\perp =0$,

$$
n^{*\mu} n^{*\nu} D_{\mu\nu}^{LC} (x) =
{2 \pi^{-D/2} \Gamma (D/2) \over 4 - D}
{(xn^*)^2 \over (- x^2 + i \epsilon)^{D/2}}.
\eqno(3.1)
$$

The calculation of the Wilson Loop under consideration at ${\cal O}(g^4)$
in 1 + (D - 1) dimensions, using light--cone gauge,
has been performed in ref. [10]. Here we shall report those results
and then perform their
Laurent expansion around  D=2, the value we are interested in.

One might wonder why dimensional regularization should
be introduced at all, as one might presume that single
graph contributions are likely to be finite in this gauge.
On the other hand, while remaining strictly at D=2, no
self--interaction should be present.

We shall discuss this point of view at the end of the paper.
For the time being let us recall that, when $D \not = 2$,
``transverse" vector components are turned on and, although their
contribution is expected to be ${\cal O}(D - 2)$, it can compete
with singularities arising from loop corrections.
This is indeed what happens in the self--energy calculation,
as will be soon apparent.

The calculation ${\cal O}(g^2)$ is easily performed and the result
exactly coincides with eq. (2.2), for any value of D.
At ${\cal O}(g^4)$ we again confine ourselves to the  ``maximally
non--abelian" contributions, without losing information. The
self--energy graph now gives

$$
W_{LC}^{(2; se)} = \Bigl({\hat g \over \pi \mu}\Bigr)^4 {C_F C_A \over
16}
{\cal A} \Bigl\{{4 \over (4 - D)^4 (D - 3)} \Bigl[{\Gamma^2 (3 - D/2) \Gamma
(D - 3) \over \Gamma (5 - D)} -
$$
$$
- \Gamma^2 (D/2 - 1)\Bigr] + {\Gamma^2
(D/2 - 1) \over (4 - D)^3 (D-3)} \Bigl[3 - {3D - 2 \over 4(D - 1)} -
{D - 2 \over D - 3}\Bigr]\Bigr\}. \eqno(3.2)
$$

Its limit at D = 2

$$
W_{LC}^{(2; se)} (D = 2) = ({g^2 \over 4 \pi})^2 C_F C_A (n^* n)^2
\eqno (3.3)
$$

is finite, but it does not vanish, as one might have
naively expected.

Similarly the contribution from the ``cross" graphs

$$
W_{LC}^{(2 ; cr)} = - ({\hat g \over \pi \mu})^4
{C_F C_A \over 16} {\Gamma^2 (D/2 - 1) \over
(D - 4)^4} \{2 {\cal A} {D - 2 \over D - 3} +
8{\cal B} [1 - 2 {\Gamma^2 (3 - D/2) \over
\Gamma (5 - D)}]\} \eqno(3.4)
$$

leads to a finite, non vanishing, result in the
limit D=2

$$
W_{LC}^{(2;cr)} (D = 2) = ({g^2 \over 4\pi})^2 C_F C_A
(n^* n)^2 {\pi^2 \over 3}. \eqno(3.5)
$$

Summing eq. (3.3) and (3.5) we exactly recover eq. (2.12).

As a matter of fact the contribution due to graphs
with three ``gluon" lines [10]

$$
W_{LC}^{(2 ; 3g)} = \Omega \Bigl\{\Gamma (D/2 - 2)\Gamma (3 - D/2) +
$$
$$
+ {\Gamma^2 (3 - D/2) \over \Gamma
(5 - D)} {6D - 28 \over (D - 2) (D - 4)} - {2 \over \Gamma
(2 - D/2)} S_1 (D) - (4 - D) \Gamma (3 - D/2) S_2 (D) \Bigr\}
\eqno(3.6)
$$

where

$$
\Omega = {2g^4 C_F C_A \over (2 \pi)^D} (2 nn^*)^{4-D}
e^{-i \pi D \over 2} cos ({\pi D \over 2}) {\Gamma (D - 4) \over
(D - 4)^2}, \eqno(3.7)
$$

$$
S_1 (D) = \sum^\infty_{n=0} {\Gamma(n +2 - D/2) \over
(n + 3 - D/2)n!} \bigl[ \psi(n + D/2) - \psi (n + 3 - D/2) +
$$
$$
+ {2 \over (n + D/2 - 1) (n + D/2)} + {1 \over
n + 3 - D/2} - {\Gamma (n + D/2 - 1) \Gamma
(5 - D) \over \Gamma (4 + n - D/2)}\bigr] \eqno(3.8)
$$

and

$$
S_2 (D) = \sum^\infty_{n=0} {\Gamma(n+3 - D/2) \over
\Gamma(n+6 - D)} \bigl[\Gamma (D/2 - 2) ({\Gamma (n+5 - D) \over
\Gamma (n+3 - D/2)} - {\Gamma(n+2) \over \Gamma(n+D/2)}) +
$$
$$
+ 2 {\Gamma(n+1) \Gamma (D/2) \over \Gamma(n+1+D/2)} +
{\Gamma(n+5 - D) \Gamma (D/2 - 1) \over \Gamma (n+4 - D/2)}
- {\Gamma(n+1) \Gamma (3 - D/2) \over \Gamma(n + 4 - D/2)}\bigr]
\eqno(3.9)
$$
vanishes when D =2.

As a consequence the same finite result for the Wilson loop
${\cal O} (g^4)$ at D=2 is obtained both in Feynman and in light--cone
gauges. However non--abelian terms are definitely present; the
theory cannot be considered a free one in quantum loop calculations
at D=2, in spite of the quadratic nature of its classical
lagrangian density in light--cone gauge. From a practical view
point,
in this fully interacting theory, the hope of getting solutions, when
quarks are included, e.g. for the mesonic spectrum, in analogy
with 't Hooft's treatment, seems to us remote.
\vfill\eject
\vskip 0.3truecm
{\bf 4. The 't Hooft approach}
\vskip 0.3truecm
In this section we stick in 1+1 dimensions. If we
interprete $x^-$ as time direction, the field $A_-$ is
not an independent dynamical variable and just provides a non--local
force of Coulomb type between fermions. In momentum space it
can be described by the ``exchange" $k_+^{-2}$ [4].

Owing to its singular IR behaviour, this expression is not a
tempered distribution; however it can be Fourier transformed after
an analytical regularization

$$
D_{--} (x) = {i \over (2 \pi)^2} \int e^{ikx} d^2k |k_+|^
{-2\lambda}\Bigl\vert_{\lambda = 1}.  \eqno(4.1)
$$

Alternatively the same result can be obtained by
interpreting the square as (minus) the derivative
of the Cauchy principal value (CPV) distribution

$$
D_{--} (x) = -{i \over (2 \pi)^2} \int e^{ikx} d^2k
{\partial \over \partial k_+} [CPV ( {1 \over k_+})] =
- {i \over 2} |x_-| \delta (x_+). \eqno(4.2)
$$

It is straightforward to check that, by inserting eq. (4.2) in
our Wilson loop, the result (2.2) at ${\cal O}(g^2)$  is recovered.

At ${\cal O}(g^4)$ in 1+1 dimensions, the only ``a priori" surviving
non--abelian contribution, which is due to ``cross" graphs,
vanishes using eq. (4.2). Henceforth no $C_F C_A$ term appears, in
agreement with abelian exponentiation, but at variance with the
result obtained (after regularization!) in Feynman gauge. On
the other hand no fully consistent vector loop calculation
would be feasible in 1+(D-1) dimensions, using a CPV prescription
or introducing IR cutoffs [16].

If we perform instead an equal time canonical quantization,
starting from the lagrangian density

$$
L = {1 \over 2} F^a_{+-} F^a_{+-} + \lambda^a nA^a,
\eqno(4.3)
$$
$\lambda^a$ being Lagrange multipliers, by imposing the equal time
commutation relations

$$
[A^a_1 (t, x), F^b_{01} (t,y)] = i \delta (x - y) \delta^{ab},
\eqno(4.4)
$$
we recover for the vector propagator exactly the ML prescription
restricted at D=2.

In this context the equation for the Lagrange multipliers
$$
n \partial \lambda^a = 0
\eqno(4.5)
$$
is to be interpreted as a true equation of motion and the
fields $\lambda^a$ provide propagating degrees of freedom,
although of a ``ghost" type [14].

The potentials $A_-^a$ have the momentum decomposition

$$
\tilde A_-^a (k) = u^a \delta' (k_+) + v^a \delta (k_+),
\eqno(4.6)
$$

$\tilde \lambda^a$ (k) being proportional to $u^a$:
$\tilde \lambda^a = k_- u^a$.

The canonical algebra (4.4) induces on $u^a$ and $v^a$
the algebra
$$
[v^a_\pm (k_-), u^b_\mp (q_-)] = \pm \delta (k_- - q_-)
\delta^{ab}, \eqno(4.7)
$$

$v^a_\pm$ and $u^a_\pm$ being defined as

$$
v^a(k_-) = \theta (k_-) v^a_+ (k_-) + \theta
(- k_-) v^a_a (- k_-),
$$

$$
u^a (k_-) = \theta (k_-) u^a_+ (k_-) - \theta (- k_-)
u^a_- (- k_-),
$$

all others commutators vanishing.

This algebra eventually produces the propagator (3.1) for D=2.
No wonder then that we recover from the ``cross" graphs
eq. (3.5), whereas, in strictly 1+1 dimensions, neither self--energy
corrections nor graphs with three vector lines should be considered.

The result we obtain in this third scenario neither coincides
with the one in Feynman gauge (the limit $D \rightarrow 2$
being ``discontinuous") , as we have neglected the non--vanishing
self--energy correction, nor obeys abelian exponentiation as in
't Hooft's approach, the reason being rooted in a different
content of the degrees of freedom (the fields $\lambda^a$).
Although perhaps more satisfactory from a mathematical
view point [7], it looks in our opinion less coherent and its
``physical" interpretation looks somewhat obscure.

\vskip 1truecm
\vfill\eject
{\bf References}
\vskip 0.3truecm
\item{[1]} E. Witten, Comm. Math. Phys. \underbar{141}
(1991) 153.

\item{[2]} E. Witten, J. of Geom. and Phys. \underbar{9} (1992) 303;

M. Blau and G. Thompson, Int. J. Mod. Phys.
\underbar{A7} (1992) 3781.

\item{[3]} M. Blau and G. Thompson, Lectures on 2d Gauge
Theories. Preprint IC/93/356, hep -- th/9310144 and references therein.

\item{[4]} G. 't Hooft, Nucl. Phys. \underbar{B75}
(1974) 461.

\item{[5]} M. Cavicchi, P. di Vecchia, I. Pesando, Mod. Phys.
Lett. \underbar{A8} (1993) 2427.

\item{[6]} C.G. Callan, N.Coote and D.J. Gross, Phys.Rev.
\underbar{D 13} (1976) 1649.

\item{[7]} T.T. Wu, Phys. Lett. \underbar{71B}
(1977) 142.

\item{[8]} N.J. Bee, P.J. Stopford and B.R. Webber, Phys. Lett.
\underbar{76B}, 315 (1978).

\item{[9]} I.A. Korchemskaya and G.P. Korchemsky,
Phys. Lett. \underbar{B 287} (1992) 169.

\item{[10]}A. Bassetto, I.A. Korchemskaya, G.P. Korchemsky and G.
Nardelli, Nucl. Phys. \underbar{B 408} (1993) 62.

\item{[11]} J.G.M. Gatheral, Phys. Lett. \underbar{B133} (1983)
90;

J. Frenkel and J. C.Taylor, Nucl. Phys. \underbar{B246}
(1984) 231.

\item{[12]} A. Bassetto, G. Nardelli and R. Soldati, Yang--Mills
theories in algebraic non covariant gauges, (World Scientific,
Singapore, 1991).

\item{[13]}S. Mandelstam, Nucl. Phys. \underbar{B213}
(1983) 149;

G. Leibbrandt, Phys. Rev. \underbar{D29} (1984) 1699.

\item{[14]} A.Bassetto, M. Dalbosco, I. Lazzizzera and R. Soldati,
Phys. Rev. \underbar{D31}(1985) 2012.

\item{[15]} A. Bassetto, Phys. Rev. \underbar{D46}
(1992) 3676.

\item{[16]} D.M. Capper, J.J. Dulwich and M.J. Litvak,
Nucl. Phys. \underbar{B241} (1984) 463.

\bye